\documentclass[prl,twocolumn,superscriptaddress]{revtex4}
\usepackage{amsmath}

\begin{document}

\title{A simple theory for the dynamics of mean-field-like models of 
glass-forming fluids}
\author{Grzegorz Szamel}
\affiliation{Department of Chemistry, 
Colorado State University, Fort Collins, CO 80523}

\date{\today}

\begin{abstract}
We propose a simple theory for the dynamics of model
glass-forming fluids, which should be solvable using a mean-field-like approach.
The theory is based on transparent physical assumptions, which can be
tested in computer simulations. The theory predicts an
ergodicity-breaking transition that is identical to the so-called dynamic
transition predicted within the replica approach. Thus, it can provide the
missing dynamic component of the random first order transition framework.
In the large-dimensional limit the theory reproduces the result of a recent exact
calculation of Maimbourg \textit{et al.} [PRL \textbf{116}, 015902 (2016)]. Our approach
provides an alternative, physically motivated derivation of this result.
\end{abstract} 

\maketitle
In the last decade, the static component of the general theoretical framework known 
as the random first order transition (RFOT) theory \cite{KWPRA1987,KTPRB1987,KWPRB1987}
was developed into a consistent, albeit arguably mean-field-like, static description of 
the glass transition \cite{ParisiZamponiRMP,BerthierBiroliRMP,MezardParisiWolLubbook}.
This has been achieved by a combination of the generalization 
of the replica approach to amorphous systems without quenched disorder 
\cite{Monasson} 
with the theoretical apparatus of the liquid state theory \cite{HansenMcDonald}. 
More recently, it has been realized that in the limit of large spatial dimension
the mean-field approach becomes exact and the complete theory of
the glass (and jamming) transition in the infinite-dimensional hard-sphere system
has been worked out \cite{largedCKPUZNC,largedCKPUZAR}.

On the other hand, the dynamic component of the RFOT approach has not been advanced 
to the same degree. This is a bit disappointing
since the glass transition, as observed either in a laboratory or a computer experiment,
manifests itself most clearly through the enormous slowing down of the dynamics. 
While within the original $p$-spin version of the RFOT approach \cite{KTPRB1987} 
(and within the very simplified hard-sphere calculation of Ref. \cite{KWPRA1987}) 
static and dynamic components of the
theory were fully consistent, there is no finite-dimensional dynamic
theory consistent with the advanced version of the replica-based static approach.
In contrast,  in the limit of large spatial dimension the dynamics of the hard-sphere 
system has been solved \cite{MaimbourgPRL}. More 
precisely, the many-body problem has been reduced to a one-dimensional 
stochastic equation with colored noise, which is determined self-consistently. 
At high density, this equation predicts an ergodicity-breaking
transition which is fully consistent with the so-called dynamic transition
predicted by the large-dimensional replica calculation. Thus, the calculation
of Maimbourg \textit{et al.} suggests that it should be possible to come up
with dynamical theories that agree with static approaches. Unfortunately, the 
physical content of the Maimbourg \textit{et al.} result 
is a bit obscured by a rather long derivation. 

One should note at this point that until the formulation of the static replica approach,
the mode-coupling theory of the glass transition \cite{Goetzebook} was the most
successful quantitative description of the glass transition 
and for a long time was considered to be the dynamic mean-field theory 
of this transition (indeed, it was featured in this role in the 
original RFOT theory papers). However, it was discovered in the last decade 
\cite{SchmidSchillingPRE,IkedaMiyazakiPRL2010} that in the large-dimensional limit 
the ergodicity-breaking transition predicted by the mode-coupling theory is 
different from the dynamic transition of the replica approach 
(which becomes exact in the large-dimensional limit).

The somewhat uncertain status of the mode-coupling theory
resulting from its inadequacy in the large-dimensional limit \cite{JPBJCCM} 
is disappointing in view of the fact that attempts to go beyond a
mean-field-like description of the dynamics usually start
from the mode-coupling theory \cite{KSS,SFH2013,Rizzo2014}. Parenthetically,
we should note that most attempts to go beyond a mean-field-like static description of 
the glass transition use lattice-based effective Hamiltonians rather than 
particle-based models \cite{CBTT2011,BCTT2014}.

Our goal in this Letter is to present a simple theory for glassy dynamics, 
which predicts ergodicity-breaking transitions consistent with dynamic
transitions predicted by the replica approach. This theory can become the
missing dynamic component of the RFOT approach. The theory is based on
transparent physical assumptions. The most important assumption is that there
should be no ``loops'' in the dynamics (this notion and its consequences are 
discussed in the following). In principle, this assumption 
makes the theory applicable only to mean-field-like models of glass-forming fluids. 
In particular, we show that the large-dimensional limit of our theory coincides with 
the exact result derived by Maimbourg \textit{et al.} In finite dimensions the 
situation is a bit more complex. Additional assumptions of the theory,
the most prominent one being the Gaussian character of the single-particle
motion, make it only an approximate description of the dynamics of finite-dimensional
mean-field-like models. On the other hand, analogous assumptions are used
in the replica theory description of finite-dimensional
mean-field-like models \cite{ParisiZamponiRMP} (although, in principle, 
one can avoid these assumptions using the cavity approach 
of M\'ezard \textit{et al.} \cite{MPTZ}). The result is that the ergodicity-breaking
transition predicted by our theory for mean-field-like models in finite dimensions 
is the same as the dynamic transition predicted by the replica description.

To make our considerations more specific, we use the Mari-Krzakala-Kurchan 
model \cite{MKK}. We consider $M$ particles in $d$ dimensional space. Any given 
particle interacts, \textit{via} a spherically symmetric
potential $V(r)$, only with $N$ other particles, with $N\ll M$. The 
network specifying inter-particle interactions forms a quenched random tree-like graph. 
This graph does not have ``loops'' and thus dynamical events in which two 
interacting particles interact with the same third particle are 
absent. This fact makes the model solvable \textit{via} a mean-field-like approach 
(at least in principle). In particular, for this model the pair correlation
function is equal to the Boltzmann factor, $g(r) = e^{-\beta V(r)}$, with $\beta=1/T$
(we use a system of units in which $k_B=1$). 
At the initial time the particles are distributed according to 
the canonical ensemble. They evolve with Brownian dynamics; thus, each particle 
is subjected to the inter-particle interaction and a random Gaussian
white noise.
We note that in the large-dimensional limit the present model system and any system 
with short-range interactions are identical. This is due to the fact that in large
dimensions the probability of ``loops'' is vanishingly small for geometric reasons.

The first (although not the most important) assumption of our approach 
is concerned with the description of the single-particle
motion. Specifically, we assume that the motion of one selected (tagged) particle
in the fluid of interacting particles can be described by the following generalized 
Langevin equation with Gaussian noise,
\begin{eqnarray}\label{taggedeom}
\gamma \dot{\mathbf{r}}_1(t) =  - \int_0^t M^{\text{irr}}(t-t') 
\dot{\mathbf{r}}_1(t') dt'
+ \boldsymbol{\eta}_1(t) + \boldsymbol{\xi}_1(t).
\end{eqnarray}
Here, $\gamma$  is the friction coefficient of an isolated particle and
$M^{\text{irr}}(t)$ is the irreducible memory function describing the average
response of the other particles of the fluid to the motion of the tagged
particle. The memory function is essentially a friction
kernel, \textit{i.e.} the first term at the right-hand-side of Eq. (\ref{taggedeom})
is the internal friction force experienced by the tagged particle due to the 
presence of other particles. Next, $\boldsymbol{\eta}_i(t)$ in Eq. (\ref{taggedeom}) 
is a Gaussian colored noise describing the  
fluctuating force acting on particle $i$ originating from the presence
of the other particles. In equilibrium, the noise should be related to the
memory function by a fluctuation-dissipation relation,
\begin{eqnarray}\label{fdt}
\left<\boldsymbol{\eta}_i(t) \boldsymbol{\eta}_j(t')\right> =  
T \delta_{ij}  \boldsymbol{I} M^{\text{irr}}(t-t'),
\end{eqnarray}
where $\boldsymbol{I}$ is the unit tensor. 
Finally, $\boldsymbol{\xi}_i(t)$ in Eq. (\ref{taggedeom}) is a  
Gaussian white noise acting on the particle $i$, with autocorrelation
function $\left<\boldsymbol{\xi}_i(t) \boldsymbol{\xi}_j(t')\right> = 
2\gamma T \delta_{ij} \boldsymbol{I} \delta(t-t')$.

The generalized Langevin equation (\ref{taggedeom}) can be justified as follows.
Using the standard projection operator considerations one can derive an exact but 
formal equation for the time evolution of the tagged density auto-correlation 
function, which involves a wave-vector dependent irreducible memory function.
From the evolution equation one can derive an equation of motion for the tagged 
particle's mean-square displacement. The latter equation involves the zero-wave-vector 
limit of the irreducible memory function. \textit{If} the particle's motion can be 
described by a Gaussian stochastic process, then one can deduce Eq. (\ref{taggedeom})
from the time evolution of the tagged particle's mean-square displacement. It 
is possible that an exact but formal equation similar to Eq. (\ref{taggedeom}) 
could be derived, but with some kind of a position dependent irreducible memory function
and non-Gaussian noise. 


The second, somewhat more technical, assumption of our approach 
is concerned with the irreducible memory function. 
One can derive an exact but formal expression for this function as an autocorrelation
function of the total force acting on the tagged particle evolving with 
the so-called irreducible dynamics \cite{CHess},
\begin{eqnarray}\label{mf0}
M^{\text{irr}}(t) =  \beta
\left< \hat{\mathbf{k}}\cdot\mathbf{F}_1 e^{\Omega^{\text{irr}}t} 
\hat{\mathbf{k}}\cdot\mathbf{F}_1 \right>.
\end{eqnarray}
Here $\hat{\mathbf{k}}$ is a unit vector, 
$\mathbf{F}_1 = \sum_{i>1} \mathbf{F}_{1i}$ is the total force acting
on the tagged particle (the summation extends over the particles that particle
$1$ interacts with) and $\Omega^{\text{irr}}$ is the irreducible evolution operator
\cite{CHess} (note that our definition of the memory function
includes an additional factor $\beta$  compared with the definition 
of Maimbourg \textit{et al.}). 
Here, following Maimbourg \textit{et al.}, we will assume that 
the memory function
can be obtained from the \textit{pair} force evolving with the standard dynamics,
\begin{eqnarray}\label{mf}
M^{\text{irr}}(t) & \approx & 
\beta \sum_{i>1} 
\left< \hat{\mathbf{k}}\cdot\mathbf{F}_{1i}e^{\Omega t}
\hat{\mathbf{k}}\cdot\mathbf{F}_{1i} \right> 
\nonumber \\ &=&
\beta \sum_{i>1} 
\left< \hat{\mathbf{k}}\cdot\mathbf{F}_{1i}(t) 
\hat{\mathbf{k}}\cdot\mathbf{F}_{1i}(0) \right>, 
\end{eqnarray}
where $\Omega$ is the standard 
Smoluchowski operator describing the motion of interacting Brownian particles. 

The validity of assumption (\ref{mf}) can be checked almost directly. 
One could evaluate the right-hand-side of Eq. (\ref{mf}) in a Brownian dynamics
computer simulation and then calculate 
the time-dependence of the mean-squared displacement from this approximate
irreducible memory function. This calculated approximate
mean-square displacement could then be compared to the mean-squared displacement
measured in the same simulation. 

It is clear that in general Eq. (\ref{mf}) is only an approximation. For example,
for the finite-dimensional model with short-range interactions even the initial 
($t=0$) values of the exact memory function (\ref{mf0}) and the approximation (\ref{mf})
are different. However, for a Mari-Krzakala-Kurchan model, due to the 
loop-less structure of the inter-particle interactions network, the initial
values of these functions are the same. It follows from Maimbourg
\textit{et al.} that this is also true in the large-dimensional limit.

The second assumption, Eq. (\ref{mf}), suggests that one
can evaluate the memory function by calculating the force between two selected
particles, evolving with the standard dynamics, given that at the initial time
these two particles were distributed according to probability distribution
$\propto \hat{\mathbf{k}}\cdot\mathbf{F}_{12} g(r_{12})$, where 
$g(r_{12})$ is the equilibrium pair correlation function.

Let us now consider the two-particle dynamics. The force acting on one of these
particles consists of the following parts. The first part is
the force due to the second particle. The second part describes the
interaction with the other particles of the fluid, \textit{i.e.} 
particles different from the interacting pair of particles. The latter force has 
the structure similar to the force acting on one particle, Eq. (\ref{taggedeom}).
It consists of an 
average part and fluctuating part. For a Mari-Krzakala-Kurchan model 
(and for any model in the large-dimensional limit) the force acting on the first 
particle and originating from other particles of the fluid should be independent of the 
state of the second particle (and \textit{vice versa}). This is due to the fact that in
these models the particles that interact with the first particle (and the particles
that interact with these particles \textit{etc.}) do not interact with the second 
particle. 
We have to emphasize here that the argument described in this paragraph
relies on our \textit{most important} assumption that the structure of the inter-particle
interactions is such that ``loops'' are absent both in the statics and in the dynamics. 

The argument formulated above leads us to assume the following equations
of motion for the dynamics of the interacting pair particles,
\begin{eqnarray}\label{paireom1a}
\gamma \dot{\mathbf{r}}_1(t) &=&  \mathbf{F}(\mathbf{r}_{12}(t))
\\ \nonumber && - \int_0^t M^{\text{irr}}(t-t') \dot{\mathbf{r}}_1(t') dt'
+ \boldsymbol{\eta}_1(t) + \boldsymbol{\xi}_1(t)
\\ \label{paireom1b}
\gamma \dot{\mathbf{r}}_2(t) &=&  -\mathbf{F}(\mathbf{r}_{12}(t))
\\ \nonumber && - \int_0^t M^{\text{irr}}(t-t') \dot{\mathbf{r}}_2(t') dt'
+ \boldsymbol{\eta}_2(t) + \boldsymbol{\xi}_2(t).
\end{eqnarray}
Obviously, to calculate the inter-particle force we only need the relative
position, $\mathbf{r} = \mathbf{r}_1-\mathbf{r}_2$.
The equation of motion for the relative position can be easily obtained
from Eqs. (\ref{paireom1a}-\ref{paireom1b}),
\begin{eqnarray}\label{paireom2}
\tilde{\gamma}\dot{\mathbf{r}}(t) = \mathbf{F}(\mathbf{r}(t))
- \int_0^t \tilde{M}(t-t') \dot{\mathbf{r}}(t') dt'
+ \boldsymbol{\eta}(t) + \boldsymbol{\xi}(t)
\end{eqnarray}
where $\tilde{\gamma} = \frac{1}{2}\gamma$, 
$\tilde{M}(t) = \frac{1}{2}M^{\text{irr}}(t)$, $\boldsymbol{\eta}(t) = 
\frac{1}{2}\left(\boldsymbol{\eta}_1(t) - \boldsymbol{\eta}_2(t)\right)$,
$\boldsymbol{\xi}(t) = 
\frac{1}{2}\left(\boldsymbol{\xi}_1(t) - \boldsymbol{\xi}_2(t)\right)$ and thus
$\left<\boldsymbol{\xi}(t) \boldsymbol{\xi}(t')\right>= 
2 \tilde{\gamma} T \boldsymbol{I} \delta(t-t')$ and 
$\left<\boldsymbol{\eta}(t) \boldsymbol{\eta}(t')\right> =  
\boldsymbol{I} T \tilde{M}(t-t')$. We note that Eq. (\ref{paireom2}) 
is to be solved with the initial condition distributed according to  
$P(\mathbf{r}) = n  \hat{\mathbf{k}}\cdot\mathbf{F}(\mathbf{r}) g(r)$ where
$n=N/V$ is the number density of particles interacting with the tagged
particle and $g(r)= e^{-\beta V(r)}$ is the pair correlation function for
the Mari-Krzakala-Kurchan model. Then, the memory function is given by
\begin{eqnarray}\label{mf2}
\tilde{M}(t) = \frac{1}{2}M^{\text{irr}}(t) = 
\frac{n\beta}{2} \int d\mathbf{r} \, \hat{\mathbf{k}}\cdot\mathbf{F}(\mathbf{r}(t))\,
\hat{\mathbf{k}}\cdot\mathbf{F}(\mathbf{r}) g(r)
\end{eqnarray}
where $\mathbf{r}(t=0)=\mathbf{r}$.

We note that even though the noise $\boldsymbol{\eta}(t)$ 
is Gaussian (since it is defined as a difference of two Gaussian noises), 
the process describing the pair dynamics, \textit{i.e.} $\mathbf{r}(t)$, 
is not Gaussian. This is due to the fact that the relation between these two processes,
\textit{i.e.} Eq. (\ref{paireom2}), is non-linear. 

In principle, the self-consistent solution of Eq. (\ref{paireom2}) determines the 
irreducible memory function. At present, an analytic solution of this
equation is not available.
We can, however, show that at high enough density Eqs. 
(\ref{paireom2}-\ref{mf2}) predict an ergodicity-breaking transition.
To this end we can use the argument similar to that used by Maimbourg \textit{et al.}
(see Sec. V.A.1 of the supplementary information for Ref. \cite{MaimbourgPRL}).

We assume that at high enough density or low enough temperature a 
two-step relaxation process sets up, with both the mean-square displacement and the
irreducible memory function being
temporarily arrested around their respective plateau values. For times in the
plateau region, both parts of the noise $\boldsymbol{\eta}(t)$ and the corresponding
part of the memory function can be treated as adiabatically slow. 
It can be showed (see the Appendix)
that the plateau value of the memory function, $\tilde{M}_{\text{EA}}$, 
can be expressed as follows,
\begin{eqnarray}\label{mfplateau}
\tilde{M}_{\text{EA}} = 
\frac{n\beta}{2} \int d\mathbf{s} P_{\text{slow}}(\mathbf{s})
\left<\hat{\mathbf{k}}\cdot\mathbf{F}(\mathbf{r})\right>_{\mathbf{s}}^2
\end{eqnarray}
where the distribution of the slow variable $\mathbf{s}$ 
(which is a linear combination of the slow part of the noise and the initial condition) 
reads 
\begin{eqnarray}\label{Pslow}
P_{\text{slow}}(\mathbf{s}) = 
\int \frac{d\mathbf{r}}{(2 \pi T\tilde{M}_{\text{EA}})^{d/2}}
e^{-\beta V(r)-\frac{\left(\mathbf{s}-\tilde{M}_{\text{EA}}\mathbf{r}\right)^2}
{2T\tilde{M}_{\text{EA}}}}
\end{eqnarray}
and the conditional average for a given value of the slow variable is defined as
\begin{eqnarray}\label{partialeq}
\left< f(\mathbf{r}) \right>_{\mathbf{s}}
= \frac{\int d\mathbf{r}
e^{-\beta 
\left(V(r)+\tilde{M}_{\text{EA}}\mathbf{r}^2/2-\mathbf{s}\cdot\mathbf{r}\right)} 
f(\mathbf{r})}
{\int d\mathbf{r}
e^{-\beta 
\left(V(r)+\tilde{M}_{\text{EA}}\mathbf{r}^2/2-\mathbf{s}\cdot\mathbf{r}\right)}}.
\end{eqnarray}
We should note that the specific form of Eqs. (\ref{Pslow}-\ref{partialeq})
follows from the assumed Gaussian character of the noise.

Eq. (\ref{mfplateau}) is a self-consistent equation for the plateau value
of the memory function, $\tilde{M}_{\text{EA}}$. A non-zero solution of this 
equation signals breaking of the ergodicity. It can be showed (see the Appendix)
that Eq. (\ref{mfplateau}) is equivalent to the equation determining the dynamic
transition within the replica approach \cite{CJPZPNAS2014}. One should recall that
the latter equation was derived using an assumption of a Gaussian shape of
the ``cage'', which corresponds to our assumption of the Gaussian character
of the single-particle motion and of the noise. 

Now, let us show that in the large-dimensional limit our theory is
identical to the exact result of Maimbourg \textit{et al.}
First, we should recognize the fact that in the large-dimensional limit
the relative motion of two particles that are interacting at the initial time
(which is the process described by Eq. (\ref{paireom2}) with the initial
condition distributed according to  
$P(\mathbf{r}) = n  \hat{\mathbf{k}}\cdot\mathbf{F}(\mathbf{r}) g(r)$) 
proceeds predominantly along the original direction of the relative
coordinate, \textit{i.e.} along $\mathbf{r}(t=0)$. In other words, 
in the large-dimensional limit $\partial_t \hat{\mathbf{r}}(t)$ can be neglected. 
This allows us to focus on the equation of motion for 
the inter-particle distance $r(t)\equiv |\mathbf{r}(t)|$.
Using It\^{o}'s convention we obtain,
\begin{eqnarray}\label{diseom1}
\tilde{\gamma} \partial_t r(t) &=&  
F(r(t))
- \hat{\mathbf{r}}(t)\cdot\int_0^t \tilde{M}(t-t') \dot{\mathbf{r}}(t') dt'
\nonumber \\ && 
+ T \frac{d-1}{r(t)}
+ \hat{\mathbf{r}}(t)\cdot\boldsymbol{\eta}(t) 
+ \hat{\mathbf{r}}(t)\cdot\boldsymbol{\xi}(t),
\end{eqnarray}
where $F(r(t))=\hat{\mathbf{r}}(t)\cdot\mathbf{F}(\mathbf{r}(t))$ and
the first term in the second line originates from It\^{o}'s lemma.
Since $\hat{\mathbf{r}}(t)\approx \hat{\mathbf{r}}(t')$, we can express 
the second term at the right-hand-side of Eq. (\ref{diseom1}) in terms of 
$r(t')$. Next, we use the scaling relationships introduced by Maimbourg \textit{et al.},
which can also be deduced from the large-dimensional limit of Eq. (\ref{mfplateau}):
$r(t) = \sigma(1+h(t)/d)$, $\hat{\gamma} = \tilde{\gamma}\sigma^2/d^2 
= \gamma\sigma^2/(2d^2)$, $F(h) = \frac{\sigma}{d} F(r)$ and 
$M(t) = \frac{\sigma^2}{d^2} \tilde{M}(t)$, with $\sigma$ being the particle diameter. 
Then, in the large-dimensional limit we 
get from Eq. (\ref{diseom1}) the following one dimensional stochastic equation 
for the ``gap'' $h(t)$,
\begin{eqnarray}\label{diseom2}
\hat{\gamma}\dot{h}(t) = 
-w'(h(t)) +  \int_0^t M(t-t') \dot{h}(t') dt'
+ \eta(t) + \xi(t)
\nonumber \\
\end{eqnarray}
where the effective force is given by $-w'(h)=F(h)+T$ and the noises $\eta$ and 
$\xi$ satisfy fluctuation-dissipation relations
$\left<\eta(t)\eta(t')\right> =  T M(t-t')$ and 
$\left<\xi(t) \xi(t')\right> = 2 \hat{\gamma} T \delta(t-t')$.
Finally, from the memory function expression (\ref{mf}) we can get an
expression for the re-scaled function $M(t)$,
\begin{eqnarray}\label{mf3}
M(t) =\frac{\beta\hat{\phi}}{2} \int dh\;  F(h) F(h(t)) e^{-\beta w(h)},
\end{eqnarray}
where $h(t=0)=h$ and $\hat{\phi}$ is the rescaled volume fraction,
$\hat{\phi} = n\mathcal{V}_d 2^d/d$, with $\mathcal{V}_d$ 
being the volume of a $d$-dimensional sphere of radius $\sigma/2$.

Eqs. (\ref{diseom2}-\ref{mf3}) are identical to the equations derived by
Maimbourg \textit{et al.} In particular, according to these equations,
there exists an ergodicity-breaking transition at $\hat{\phi}_d = 4.807$.

Let us comment on the connection
of the present simple approach and the mode-coupling theory. The most important
approximation of the latter theory is the factorization approximation in which
a four-point correlation function is replaced by a product of two-point 
correlation functions. In the present approach, the analogue of this approximation
would be neglecting the inter-particle force term, $\mathbf{F}(\mathbf{r})$, in
the stochastic equation of motion for the relative position, Eq. (\ref{paireom2}).
However, if one were just to neglect the force term in Eq. (\ref{paireom2}), 
the expression (\ref{mf2}) for the memory function would give an infinite result.
The way out is to incorporate one of the additional approximations of the
mode-coupling theory \cite{Goetzebook} and to combine discarding the force term with  
replacing the ``bare'' forces in the memory function expression (\ref{mf2}) by
renormalized forces given by the derivatives of the direct correlation function.
In the present case of the Mari-Krzakala-Kurchan model the last step amounts to
the replacement 
$\mathbf{F}(\mathbf{r}) \rightarrow T \partial_{\mathbf{r}} e^{-\beta V(r)}$ in
expression (\ref{mf2}). It can be showed that this 
procedure results in the following self-consistent 
equation for the plateau value of the memory function, 
$\tilde{M}_{\text{EA}}^{\text{mct}}$, where the superscript $\text{mct}$ indicates
a mode-coupling-like approximation,
\begin{eqnarray}\label{mfplateaumct}
\tilde{M}_{\text{EA}}^{\text{mct}} &=& 
\frac{n\beta}{2} \int \frac{d\mathbf{r}d\mathbf{r}'}
{\left(4\pi T/\tilde{M}_{\text{EA}}^{\text{mct}}\right)^{d/2}}
e^{-\frac{\left(\mathbf{r}-\mathbf{r}'\right)^2 \tilde{M}_{\text{EA}}^{\text{mct}}}
{4T}}
\nonumber \\ && \times
g(r)\hat{\mathbf{k}}\cdot\mathbf{F}(\mathbf{r})
g(r')\hat{\mathbf{k}}\cdot\mathbf{F}(\mathbf{r}').
\end{eqnarray}
By rewriting the right-hand-side of this equation 
as an integral in reciprocal space we can show that this equation is
identical to Eq. (4.6') of Kirkpatrick and Wolynes \cite{KWPRA1987} (with additional
factors of 2 in a couple of places). In the large-dimensional
limit this equation predicts an ergodicity-breaking transition at 
$\hat{\phi}_d^{\text{mct}} = \sqrt{8\pi e}= 8.265$. 
This finding agrees perfectly with the numerical
result of Ikeda and Miyazaki \cite{IkedaMiyazakiPRL2010} obtained by taking 
the large-dimensional limit of the standard mode-coupling equation for the 
non-ergodicity parameter and assuming a Gaussian form of this parameter.
We note that the mode-coupling-like version of our theory 
predicts the ergodicity-breaking transition
at a higher value of the rescaled volume fraction. This is reasonable in that
the mode-coupling-like version neglects direct interaction between the two 
particles; replacing ``bare'' forces by renormalized ones only partially
compensates for this fact.

In summary, we presented here a simple theory for the dynamics of models 
of structural glasses which should be solvable using a mean-field-like
approach. The ergodicity-breaking transition predicted by our theory
coincides with the dynamic transition of the replica approach. Thus, our theory 
provides the dynamic counterpart of the static replica approach. 

In our theory, any given particle interacts explicitly only
with one other particle of the system. Other interactions are accounted
for by a combination of a friction force and a fluctuating force. 

In the large-dimensional limit our theory reduces itself to the result
of the exact calculation of Maimbourg \textit{et al.} and thus it provides
an alternative, physically motivated derivation of this result. In finite 
dimensions it suffers from the same problem as the replica approach, \textit{i.e.}
from an additional assumption of the Gaussian character of the single-particle
dynamics. It would be very interesting to develop an analogue of a cavity
approach of M\'ezard \textit{et al.} which could possibly overcome this problem.

Our theory can shed some light onto somewhat abstract considerations of
the replica approach. In particular, if we used the potential 
of the mean force $V^{\text{mf}}(r) = -T\ln g(r)$ 
rather than the true potential in Eqs. (\ref{paireom2}-\ref{mf2}),
the resulting ergodicity-breaking transition would coincide with the dynamic transition 
predicted by a recent version of the replica theory for a standard 
finite-dimensional system \cite{MangeatZamponi}. Thus,
as far as location of the dynamic transition is concerned, the latter approach
approximates the standard hard-sphere system by a Mari-Krzakala-Kurchan system
with particles interacting via a potential of mean force. 
Conversely, this observation means that by using the potential of the 
mean force, our simple theory can be made quantitatively accurate for 
the 3-dimensional hard sphere system, at least for the location of the dynamic 
transition. 

Finally, we showed that a small modification of the present theory results in a 
mode-coupling-like approach with an additional assumption of Gaussian fluctuations.
It would be interesting to start from a more standard mode-coupling-like approach
by considering equations of motion for density fields and to develop an 
approximate theory by keeping only dynamical events which are included in the 
present approach. This might be another avenue that would allow us to avoid or 
relax the assumption of Gaussian fluctuations and thus arrive at a theory
that both provides a reasonable, albeit mean-field-like, description of the dynamics
of finite-dimensional systems and has the correct large-dimensional limit. 

I benefited from stimulating discussions with Francesco Zamponi and 
Jorge Kurchan. I also thank both of them and Elijah Flenner for comments 
on the manuscript. I gratefully acknowledge the support of NSF Grant No.~CHE 121340.

\newpage
\setcounter{section}{1}
\setcounter{equation}{0}
\numberwithin{equation}{section}
\widetext
\section*{Appendix}

\subsection{I. Equation determining the ergodicity-breaking transition}

Following Maimbourg \textit{et al.} we assume that the noise 
$\boldsymbol{\xi}(t)$, which describes the influence of other particles on
the pair of interacting particles, can be split into a fast part and a slowly evolving
part,
\begin{eqnarray}\label{noisesplit}
\boldsymbol{\eta}(t) = \boldsymbol{\eta}^f(t) + \bar{\boldsymbol{\eta}}(t)
\end{eqnarray}
with
\begin{eqnarray}\label{memsplit} 
\left<\boldsymbol{\eta}^f(t) \boldsymbol{\eta}^f(t')\right> =  
\boldsymbol{I} T \tilde{M}^f(t-t') \text{ and }
\left<\bar{\boldsymbol{\eta}}(t) \bar{\boldsymbol{\eta}}(t')\right> =  
\boldsymbol{I} T \tilde{M}^s(t-t'),
\end{eqnarray}
where $\tilde{M}^f$ evolves on the fast time scale $\tau_f$ and 
$\tilde{M}^s$ evolves on the slow time scale $\tau_s$. 

Then, we re-write the equation of motion for the relative position for times in
the plateau region, using the above formulated assumption,
\begin{eqnarray}\label{splitpair1}
\tilde{\gamma} \dot{\mathbf{r}}(t) &=&  
\mathbf{F}(\mathbf{r}(t)) 
- \tilde{M}_{\text{EA}} \mathbf{r}(t)  + \mathbf{s}
- \int_0^t \tilde{M}^f(t-t') \dot{\mathbf{r}}(t') dt'
+ \boldsymbol{\eta}^f(t) + \boldsymbol{\xi}(t),
\end{eqnarray}
where $\mathbf{s} = \tilde{M}_{\text{EA}}\mathbf{r}(0)+ \bar{\boldsymbol{\eta}}(t)$ 
and $\tilde{M}_{\text{EA}}$ is the value of the slow part of the memory function
$\tilde{M}^s$ in the plateau region.

At intermediate time scales (when $\mathbf{s}$ can be considered constant),
the relative position relaxes to an ``equilibrium'' state for a given 
value of $\mathbf{s}$. In this state $\mathbf{r}$ is distributed according
to the probability distribution 
\begin{eqnarray}\label{condeq}
P(\mathbf{r}|\mathbf{s}) = 
\frac{e^{-\beta H(\mathbf{r}|\mathbf{s})}}{Z_{\mathbf{s}}}
\end{eqnarray}
where
\begin{eqnarray}
H(\mathbf{r}|\mathbf{s}) = V(r) + \frac{1}{2} \tilde{M}_{\text{EA}} \mathbf{r}^2
-\mathbf{s}\cdot\mathbf{r}
\end{eqnarray}
and the normalization constant reads
\begin{eqnarray}
Z_{\mathbf{s}} = \int d\mathbf{r} 
\exp\left(-\beta\left(V(r) + \frac{1}{2} \tilde{M}_{\text{EA}} \mathbf{r}^2
-\mathbf{s}\cdot\mathbf{r}\right)\right).
\end{eqnarray}

Now, to calculate the value of the slow part of the memory function
we use Eq. (\ref{mf2}) of the main text. We replace $\mathbf{F}(\mathbf{r}(t))$
by the force averaged with the conditional equilibrium distribution 
(\ref{condeq}), \textit{i.e.} by 
\begin{eqnarray}\label{forcecondeq}
\left<\mathbf{F}(\mathbf{r})\right>_{\mathbf{s}}
= \frac{1}{Z_{\mathbf{s}}} \int d\mathbf{r} e^{-\beta H(\mathbf{r}|\mathbf{s})} 
\mathbf{F}(\mathbf{r}), 
\end{eqnarray}
and then average over the slowly evolving noise $\bar{\boldsymbol{\eta}}$, auxiliary
variable $\mathbf{s}$ and the initial condition, which to avoid confusion we
denote here by $\mathbf{r}_0$. Thus,
\begin{eqnarray}\label{sceqM}
\tilde{M}_{\text{EA}} &=& \frac{n\beta}{2} \int d\mathbf{r}_0 e^{-\beta V(r_0)} 
\hat{\mathbf{k}}\cdot\mathbf{F}(\mathbf{r}_0) 
\int d\mathbf{s} \int 
\frac{d\bar{\boldsymbol{\eta}}}{(2\pi T \tilde{M}_{\text{EA}})^{d/2}}
\delta(\mathbf{s} - \tilde{M}_{\text{EA}} \mathbf{r}_0 - \bar{\boldsymbol{\eta}})
e^{-\bar{\boldsymbol{\eta}}^2/2T\tilde{M}_{\text{EA}} }
\left<\hat{\mathbf{k}}\cdot\mathbf{F}(\mathbf{r})\right>_{\mathbf{s}}
\nonumber \\ &=& 
\frac{n\beta}{2} \int 
\frac{d\mathbf{s}}{(2\pi T \tilde{M}_{\text{EA}})^{d/2}}
\int d\mathbf{r}_0 e^{-\beta V(r_0)} 
\hat{\mathbf{k}}\cdot\mathbf{F}(\mathbf{r}_0)
e^{-(\mathbf{s}-\tilde{M}_{\text{EA}}\mathbf{r}_0)^2/2T\tilde{M}_{\text{EA}}} 
\left<\hat{\mathbf{k}}\cdot\mathbf{F}(\mathbf{r})\right>_{\mathbf{s}}
\nonumber \\ &=& 
\frac{n\beta}{2} \int \frac{d\mathbf{s}}{(2\pi T\tilde{M}_{\text{EA}})^{d/2}} 
e^{-\mathbf{s}^2/2T\tilde{M}_{\text{EA}}} Z_{\mathbf{s}}
\left<\hat{\mathbf{k}}\cdot\mathbf{F}(\mathbf{r})\right>_{\mathbf{s}}^2
= \frac{n\beta}{2}  \int  d\mathbf{s} P_{\text{slow}}(\mathbf{s}) 
\left<\hat{\mathbf{k}}\cdot\mathbf{F}(\mathbf{r})\right>_{\mathbf{s}}^2,
\end{eqnarray}
where $P_{\text{slow}}(\mathbf{s})$ is given by Eq. (\ref{Pslow}) of the main text.

\setcounter{section}{2}
\setcounter{equation}{0}
\subsection{II. Ergodicity-breaking transition \textit{vs.} dynamic transition
of the replica approach}

First, we re-write Eq. (\ref{mfplateau}) of the main text for the hard-sphere
system. To this end, we follow the standard procedure: we 
write $e^{-\beta V(r)}\mathbf{F}(\mathbf{r})$ as $T\partial_{\mathbf{r}}e^{-\beta V(r)}$
and then take the hard sphere limit. In this way we obtain 
the following self-consistent equation,
\begin{eqnarray}\label{sceqMhs}
\tilde{M}_{\text{EA}} = 
\frac{nT}{2d} \int \frac{d\mathbf{s}\;
e^{-\frac{\mathbf{s}^2}{2T\tilde{M}_{\text{EA}}}} }{(2\pi T\tilde{M}_{\text{EA}})^{d/2}} 
\frac{\int d\mathbf{r}_1 \delta(r_1-\sigma) 
e^{-\beta\left(\tilde{M}_{\text{EA}} \mathbf{r}_1^2/2 
-\mathbf{s}\cdot\mathbf{r}_1\right)} \int d\mathbf{r}_2 \delta(r_2-\sigma) 
e^{-\beta\left(\tilde{M}_{\text{EA}} \mathbf{r}_2^2/2 
-\mathbf{s}\cdot\mathbf{r}_2\right)} \hat{\mathbf{r}}_1\cdot \hat{\mathbf{r}}_2
}{\int d\mathbf{r}_3 \theta(r_3-\sigma) 
e^{-\beta\left(\tilde{M}_{\text{EA}} \mathbf{r}_3^2/2 
-\mathbf{s}\cdot\mathbf{r}_3
\right)}} 
\end{eqnarray}
We note that to get Eq. (\ref{sceqMhs}) we also averaged over arbitrary direction 
of unite vector $\hat{\mathbf{k}}$. 

Now, let us recall the equation which determines the dynamic transition of the
replica approach. Eq. (S20) of Supporting Information of Ref. 
\cite{CJPZPNAS2014} reads
\begin{eqnarray}\label{replicadyn1}
1= \frac{2^d \phi}{d} \frac{A}{1-m} \frac{\partial G(m,A)}{\partial A}
\end{eqnarray}
where the right-hand side is to be calculated in the limit of $m\to 1$. 
In Eq. (\ref{replicadyn1}), 
$\phi$ is the volume fraction, $\phi = n\mathcal{V}_d 2^{-d}$, with $\mathcal{V}_d$
being the volume of the sphere of radius $\sigma/2$.  
We use Appendix C of Ref. \cite{ParisiZamponiRMP} and write $G(m,A)$ as follows
\begin{eqnarray}\label{Gma1}
G(m,A) = \frac{1}{\mathcal{V}_d} \int d\mathbf{r} 
\theta(r-\sigma) \int d\mathbf{r}_1 f_{2A}^G(\mathbf{r}_1) 
\left(q_A(\mathbf{r}-\mathbf{r}_1)^{m-1} - 1\right)
\end{eqnarray}
with 
\begin{eqnarray}\label{qA1}
q_A(r) = \int d\mathbf{r}_2 f_{2A}^G(\mathbf{r}_2)
\theta(|\mathbf{r}-\mathbf{r}_2|-\sigma),
\end{eqnarray}
and $f_{A}^G$ being a Gaussian distribution,
\begin{eqnarray}\label{fAG}
f_{A}^G(\mathbf{r}) = \frac{e^{-\mathbf{r}^2/2A}}{(2\pi A)^{d/2}}.
\end{eqnarray}
The dynamic transition occurs at the lowest density at which Eq. (\ref{replicadyn1})
has a non-trivial solution for parameter $A$. This parameter is proportional to the
plateau value of the mean-square displacement, which in the dynamical approach
is inversely proportional 
to the plateau value of the memory function. The final relation 
between $A$ and $\tilde{M}_{\text{EA}}$ has the following form,
\begin{eqnarray}\label{AvsMEA}
A=\frac{T}{2\tilde{M}_{\text{EA}}}.
\end{eqnarray}

After some transformations, 
the $m\to 1$ limit of the right-hand-side of Eq. (\ref{replicadyn1}) can be 
written explicitly as follows,
\begin{eqnarray}\label{derG1}
&& \lim_{m\to 1} \frac{2^d \phi}{d} \frac{A}{1-m} \frac{\partial G(m,A)}{\partial A} 
= - \frac{n}{2d}\int d\mathbf{r} q_A^{-1}(\mathbf{r})  \int d\mathbf{r}_1  
\int d\mathbf{r}_2
\theta(|\mathbf{r}+\mathbf{r}_1|-\sigma) f_{2A}^G(\mathbf{r}_1) 
f_{2A}^G(\mathbf{r}_2-\mathbf{r})
\delta(r_2-\sigma)
\hat{\mathbf{r}}_2\cdot
(\mathbf{r}_2-\mathbf{r})
\nonumber \\ && 
+ \frac{n}{2d}
\int d\mathbf{r} q_A^{-1}(\mathbf{r})  \int d\mathbf{r}_1  \int d\mathbf{r}_2
\theta(|\mathbf{r}+\mathbf{r}_1|-\sigma) f_{2A}^G(\mathbf{r}_1) 
f_{2A}^G(\mathbf{r}_2-\mathbf{r})
\delta(r_2-\sigma)
\hat{\mathbf{r}}_2\cdot\mathbf{r}_1
\end{eqnarray}
It can be showed that the first term at the right-hand-side of Eq. (\ref{derG1}) 
vanishes. The second can be re-written as
\begin{eqnarray}\label{derG2}
\frac{nA}{d}\int d\mathbf{r} q_A^{-1}(\mathbf{r})  \int d\mathbf{r}_1  \int d\mathbf{r}_2
f_{2A}^G(\mathbf{r}_1-\mathbf{r}) \hat{\mathbf{r}}_2\cdot\hat{\mathbf{r}}_1
\delta(r_1-\sigma)
f_{2A}^G(\mathbf{r}_2-\mathbf{r})
\delta(r_2-\sigma).
\end{eqnarray}

The resulting self-consistent equation for parameter $A$ thus reads
\begin{eqnarray}\label{replicadyn2}
1 &=& \frac{nA}{d}
\int d\mathbf{r} e^{- \mathbf{r}^2/4A}
\left[\int \frac{d\mathbf{r}_1}{(4\pi A)^{d/2}}  
e^{-\mathbf{r}_1^2/4A+\mathbf{r}\cdot\mathbf{r}_1/2A} \theta(r_1-\sigma)\right]^{-1}
\nonumber \\ && \times
\int  \frac{d\mathbf{r}_1}{(4\pi A)^{d/2}}  
e^{-\mathbf{r}_1^2/4A+\mathbf{r}\cdot\mathbf{r}_1/2A}
\int  \frac{d\mathbf{r}_2}{(4\pi A)^{d/2}}  
e^{-\mathbf{r}_1^2/4A+\mathbf{r}\cdot\mathbf{r}_2/2A}
\hat{\mathbf{r}}_2\cdot\hat{\mathbf{r}}_1
\delta(r_1-\sigma)\delta(r_2-\sigma),
\end{eqnarray}
which, with identification (\ref{AvsMEA}) and a simple change of variables 
can be written as Eq. (\ref{sceqMhs}). 
Thus, the ergodicity-breaking transition of our simple dynamical theory
coincides with the dynamic transition of the replica approach.

\end{document}